\documentclass[twocolumn,prb,aps,amsmath,amssymb,longbibliography,10pt]{revtex4-1}

\usepackage{graphicx}
\usepackage{dcolumn}
\usepackage{bm}

\usepackage[%
  colorlinks=true,
  urlcolor=red,
  linkcolor=red,
  citecolor=red
]{hyperref}

\usepackage[caption=false]{subfig}
\usepackage{xcolor}

\newcommand{\bra}[1]{\ensuremath{\left\langle#1\right|}}
\newcommand{\ket}[1]{\ensuremath{\left|#1\right\rangle}}

\makeatletter
\newsavebox{\@brx}
\newcommand{\llangle}[1][]{\savebox{\@brx}{\(\m@th{#1\langle}\)}%
  \mathopen{\copy\@brx\kern-0.5\wd\@brx\usebox{\@brx}}}
\newcommand{\rrangle}[1][]{\savebox{\@brx}{\(\m@th{#1\rangle}\)}%
  \mathclose{\copy\@brx\kern-0.5\wd\@brx\usebox{\@brx}}}
\makeatother

\begin{document}

\title{A Model for Metastable Magnetism in the Hidden-Order Phase of \texorpdfstring{URu$_2$Si$_2$}{URu2Si2}}


\author{Lance Boyer}
 \email{lboyer@umd.edu}
\author{Victor M. Yakovenko}%
 \email{yakovenk@physics.umd.edu}
\affiliation{%
Condensed Matter Theory Center and Joint Quantum Institute,\\
Department of Physics, University of Maryland, College Park, Maryland 20742, USA
}%

\date{November 16, 2017}

\begin{abstract}
We propose an explanation for the experiment by Schemm \textit{et al}.,~Phys.~Rev.~B \textbf{91}, 140506 (2015) where the polar Kerr effect (PKE), indicating time-reversal symmetry (TRS) breaking, was observed in the hidden-order (HO) phase of URu$_2$Si$_2$. The PKE signal on warmup was seen only if a training magnetic field was present on cool-down. Using a Ginzburg-Landau model for a complex order parameter, we show that the system can have a metastable ferromagnetic state producing the PKE, even if the HO ground state respects TRS. We predict that a strong reversed magnetic field should reset the PKE to zero.
\end{abstract}

\maketitle

\section{Introduction} The heavy-fermion material URu$_2$Si$_2$ exhibits a second-order phase transition from paramagnetism to a puzzling hidden order (HO) at $T_{\mathrm{HO}}=17.5$~K \cite{RevModPhys.83.1301,PhysRevLett.55.2727}, where the corresponding symmetry breaking has not been definitively established. Particularly interesting is the question of whether time-reversal (TR) symmetry is preserved or broken in the HO phase. Raman spectroscopy gives evidence for the spontaneous breaking of mirror symmetries, so Kung \textit{et al}.~\cite{kung2015chirality} interpreted HO as a chirality density wave that preserves TR symmetry. However, Schemm \textit{et al}.~\cite{PhysRevB.91.140506} observed a non-zero polar Kerr effect (PKE) in the HO phase, indicating possible TR symmetry breaking \footnote{The primary focus of Ref.~\onlinecite{PhysRevB.91.140506} was on TR symmetry breaking in the superconducting phase of URu$_2$Si$_2$ below $T_c=1.5$~K, whereas our focus is on TR symmetry breaking in the HO phase, which was also studied in Ref.~\onlinecite{PhysRevB.91.140506}.}. Here, we attempt to reconcile the experimental results of Refs.~\onlinecite{kung2015chirality} and \onlinecite{PhysRevB.91.140506} within a unified theoretical framework based on an earlier model of HO developed by Haule and Kotliar in Refs.~\onlinecite{haule2009arrested, haule2010complex}.

According to Ref.~\onlinecite{PhysRevB.91.140506}, URu$_2$Si$_2$ exhibits zero PKE when cooled without an applied magnetic field, which is consistent with TR symmetry preservation in the HO phase. However, when URu$_2$Si$_2$ is cooled in a training magnetic field up to 2~T, which is then removed at low temperature, a non-zero PKE is observed on warmup in the HO phase. Apparently, the external magnetic field induces magnetism in the material, which is preserved even after the field has been removed.  Schemm \textit{et al}.~\cite{PhysRevB.91.140506} interpreted this persistent magnetism as extrinsic in origin, resulting from unspecified magnetic states due to strain or defects. While explanations due to sample inhomogeneity are possible \cite{chandra2003case,amato2004weak,PhysRevLett.87.087203,mason1990neutron}, here we advance an alternative proposition that the induced magnetism is intrinsic to HO and would occur even in a perfectly uniform sample.
 
We approach this problem from the perspective of the Haule-Kotliar model \cite{haule2009arrested, haule2010complex} characterized by a two-component complex order parameter. The real part represents chiral order consistent with the observations of Ref.~\onlinecite{kung2015chirality}, whereas the imaginary part represents magnetic order. Using a modified version of the associated free energy, we study the interplay and competition between the two components of the order parameter. We find that, when the system is cooled in a magnetic field, it may become trapped at a local minimum of the free energy, corresponding to a metastable ferromagnetic (FM) state and exhibiting the PKE. This conjecture of a metastable FM state is supported by the observation of hysteresis in direct magnetization measurements in single crystals of URu$_2$Si$_2$ cooled in zero and non-zero fields\cite{PARK1998455}.

Our proposition can be tested by applying a reversed magnetic field at low temperature. We predict that, when the reversed field exceeds a certain threshold, the system will make an irreversible transition from the metastable FM to the true HO ground state, thereby resetting the PKE (or magnetization) to zero. In contrast, an extrinsic FM would change sign in a reversed magnetic field instead of being eliminated. An experimental verification of this prediction would be a crucial test of the metastable intrinsic FM scenario and would qualitatively discriminate it from other possible explanations of the induced PKE.

\section{Haule-Kotliar Model}\label{hkmodel}
URu$_2$Si$_2$ is a body-centered tetragonal crystal, where uranium atoms are arranged in square-lattice layers perpendicular to the $c$ axis. The crystal has a four-fold rotational symmetry about the $c$ axis and four vertical mirror planes (VMP) through the $c$ axis. According to Ref.~\onlinecite{haule2009arrested}, the 5f$^2$ electrons of the uranium atoms have the ground state $\ket{A_2} = i(\ket{4,4}-\ket{4,-4})/\sqrt{2}$ and the lowest excited state $\ket{A_1} = \cos \phi (\ket{4,4}+\ket{4,-4})/\sqrt{2}+\sin \phi \ket{4,0}$, written in the angular momentum basis $\ket{J,J_z}$, where the $z$ axis is taken along the $c$ axis, and $\phi \approx 0.37\pi$. Inelastic non-resonant X-ray spectroscopy supports the conjecture that $\ket{A_1}$ and $\ket{A_2}$ are indeed the low-lying states of the system \cite{sundermann2016direct}.

A model Hamiltonian $H$ consistent with both VMP and TR symmetries can be constructed \cite{haule2010complex} using Pauli matrices $\{\sigma^x_j,\sigma^y_j,\sigma^z_j\}$ in the basis of $\ket{A_2}_j$ and $\ket{A_1}_j$ at each uranium site labeled by coordinate $j$:
\begin{align}\label{eqn:hamiltonian}
H = \sum_{\substack{\langle j,k\rangle}}[ J^x_{jk}\sigma^x_j\sigma^x_k+ J^y_{jk}\sigma^y_j\sigma^y_k] - \sum_{j}[\Delta \sigma^z_j+b\sigma^y_j].
\end{align}
Here $2\Delta = 35$ K is the energy splitting of the $A_1$ and $A_2$ states \footnote{Our definition of $\Delta$ differs by a factor of 2 from Ref.~\onlinecite{haule2010complex}. We choose $\Delta = 17.5$ K for consistency with Ref.~\onlinecite{haule2010complex}.}, $b = \mu_\mathrm{eff} B$ is the energy of interaction with an external magnetic field $B$ applied along the $c$ axis, and the amplitudes $J^{x,y}_{jk}$ describe interaction between the nearest neighboring sites $\langle j,k\rangle$. The Pauli matrices transform as $\sigma^{x,y} \rightarrow -\sigma^{x,y}$ and $\sigma^z \rightarrow \sigma^z$, upon VMP reflections because $\ket{A_2}$ is odd and $\ket{A_1}$ is even. Thus, the first two terms in $H$ are bilinear in $\sigma^{x,y}$, and the third term is linear in $\sigma^z$. Additionally, $\sigma^y \rightarrow -\sigma^y$ upon TR due to complex conjugation, so $\sigma^y$ couples linearly to the magnetic field in the last term.

At low temperature, the system described by Eq.~\eqref{eqn:hamiltonian} may undergo a phase transition that breaks VMP symmetries and results in hybridization of the even $\ket{A_1}$ and odd $\ket{A_2}$ states. It is characterized by the anomalous average
\begin{align}
\psi_j = 2 \bra{A_1}_j \rho \ket{A_2}_j =  \mathrm{Tr}[\rho (\sigma^x_j + i\sigma^y_j) ] = \psi^x_j + i\psi^y_j, 	
\end{align}
where $\rho$ is the density matrix, whereas $\psi^x_j = \langle \sigma^x_j \rangle$ and $\psi^y_j = \langle \sigma^y_j \rangle$ are the real and imaginary parts of the complex order parameter $\psi_j$. The real part represents HO and is equivalently characterized by a non-zero expectation value $\psi^x_j = -\mathrm{Tr}[\rho J_x J_y (J^2_x-J^2_y)]/8\cos\phi$ of the hexadecapolar operator \cite{haule2009arrested}, which is antisymmetric with respect to VMP reflections and symmetric with respect to TR. The associated ground state is a real superposition of $\ket{A_2}_j$ and $\ket{A_1}_j$ asymmetric with respect to VMP reflections, so it breaks chiral symmetry \cite{kung2015chirality} but preserves TR symmetry. The imaginary part of the order parameter $\psi^y_j = \mathrm{Tr}[\rho J_z]/4\cos\phi$ represents a magnetic moment along the $c$ axis and is non-zero for a complex superposition of $\ket{A_2}_j$ and $\ket{A_1}_j$. Below, we analyze the emergence of the chiral and magnetic orders using a mean-field theory.

In the mean-field approximation $\sigma^\alpha_n \sigma^\beta_m \rightarrow \psi^\alpha_n \sigma^\beta_m + \sigma^\alpha_i \psi^\beta_{m} - \psi^\alpha_{n} \psi^\beta_{m}$, the free energy at temperature $T = 1/\beta$ is given by
\begin{eqnarray}
&&F =  \sum_j \gamma(\psi^x_j\psi^y_j)^2-T\ln\left[\cosh\left(\beta\lambda_j\right)\right]-\sum_{\substack{\langle j,k\rangle\\\alpha=x,y}}J_{jk}^\alpha  \psi^\alpha_j  \psi^\alpha_k \nonumber\\
&&\lambda_j = \sqrt{\Delta^2 + \left({\sum_k J^x_{jk}\psi^x_k}\right)^2+\left({\sum_k J^y_{jk}\psi^y_k-b}\right)^2}.
 \label{eqn:hklandau}
\end{eqnarray}
Here we have introduced the additional term $\gamma(\psi^x_j\psi^y_j)^2$ with $\gamma > 0$ to discourage on-site co-existence of the chiral and magnetic orders, which is necessary to account for the first-order phase transition between HO and antiferromagnetism (AF) under pressure \cite{RevModPhys.83.1301}.

Elastic neutron scattering in the high-pressure AF phase \cite{PhysRevLett.58.1467,PhysRevB.82.060408} reveals a magnetic order that is uniform within layers, but staggered between adjacent layers, thus doubling the unit cell along the $c$ axis. A similar $c$-axis period doubling is also discussed for the HO phase, based on ARPES measurements \cite{bareille2014momentum} and the ``adiabatic continuity'' between the HO and AF phases seen in resistivity studies \cite{PhysRevLett.98.166404,PhysRevLett.105.216409}. Therefore, we take HO to be staggered, $\psi^x_n = (-1)^n\psi_{\mathrm{HO}}$, as a function of the layer number $n$, in agreement with the notion of a chirality density wave \cite{kung2015chirality}. Similarly, we decompose the magnetic order into the uniform and staggered components, $\psi^y_n = \psi_{\mathrm{FM}} + (-1)^n\psi_{\mathrm{AF}}$, representing FM and AF. Then, we rewrite Eq.~\eqref{eqn:hklandau} in terms of the three order parameters $\psi_{\mathrm{HO}}$, $\psi_{\mathrm{AF}}$, and $\psi_{\mathrm{FM}}$ coupled to the effective interaction constants $J^\alpha_{\pm} = -(4J^\alpha_{\parallel} \pm 8J^\alpha_{\bot})$, where $J^\alpha_{\parallel} <  0$ and $J^\alpha_{\bot} > 0$ are the intralayer and interlayer values of $J^\alpha_{ij}$. Positive values of the interaction amplitudes $J^x_- > J^y_- > J^y_+ > 0$ favor HO over AF over FM.

\section{Competition of Hidden Order and Antiferromagnetism} Equation~\eqref{eqn:hklandau} was used in Ref.~\onlinecite{haule2010complex} to study the interplay between HO and AF as a function of pressure in the absence of magnetic field. In this case $\psi_{\mathrm{FM}}=0$, and free energy per site $f = F/N$ ($N$ is the site count) is
\begin{eqnarray}
&&f[\psi_{\mathrm{HO}},\psi_{\mathrm{AF}}] = J^x_- \psi_{\mathrm{HO}}^2 + J^y_- \psi_{\mathrm{AF}}^2+\gamma\psi_{\mathrm{HO}}^2\psi_{\mathrm{AF}}^{2}\label{eqn:AFFreeEnergy}\\
&&-T\ln\left[\cosh\left(\beta \sqrt{\Delta^2 + \left({J^x_- \psi_{\mathrm{HO}}}\right)^2+\left({J^y_- \psi_{\mathrm{AF}}}\right)^2}\right)\right]. \nonumber
\end{eqnarray}

\begin{figure}[ht]
\includegraphics[width=0.47\textwidth]{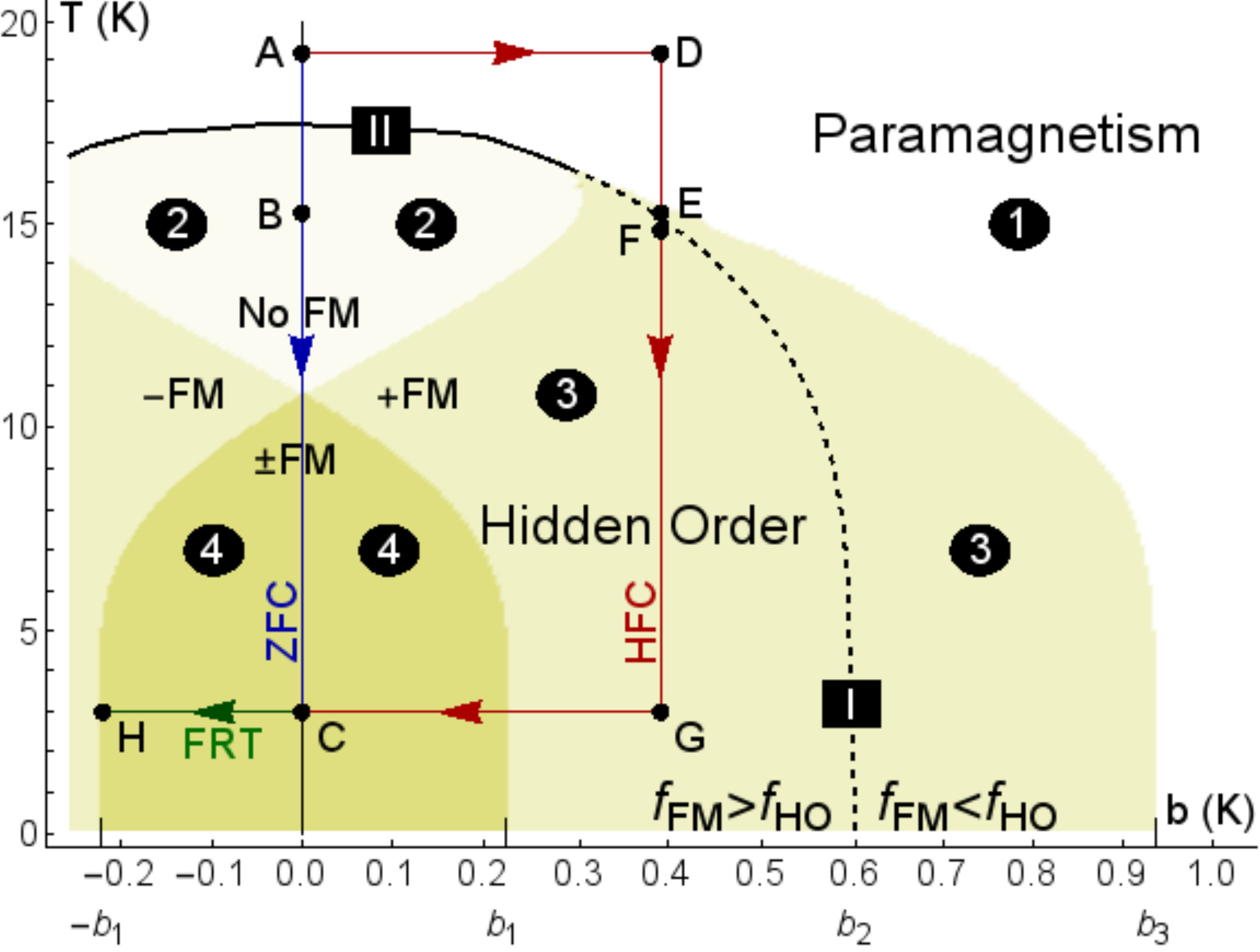}
\caption{Phase diagram for the free energy in Eq.~\eqref{eqn:IntFreeEnergy} as a function of magnetic energy $b$ and temperature $T$. The numbers in circles and the degree of shading indicate the number of minima of $f[\psi_{\mathrm{HO}},\psi_{\mathrm{FM}}]$. Every shaded domain has two degenerate HO minima with $|\psi_{\mathrm{HO}}| \ne 0$ and may have one or two FM minima with $\psi_{\mathrm{FM}} > 0$ or $\psi_{\mathrm{FM}} < 0$, as schematically indicated around $T=10$~K. The HO (FM) minima have lower energy to the left (right) of the dashed first-order transition line labeled I. The solid line labeled II represents a second-order phase transition from paramagnetism to HO. Blue, red, and green lines represent the Zero-Field Cooling (ZFC), High-Field Cooling (HFC), and Field-Reversal Test (FRT) protocols.}\label{fig:phase_diagram}
\end{figure}

Let us examine how the energy landscape given by Eq.~\eqref{eqn:AFFreeEnergy} changes with the decrease of temperature for points A, B, and C on the schematic phase diagram in Fig.~\ref{fig:phase_diagram}. In Figs.~\ref{fig:LFC}(a)-(c) we show contour plots of $f[\psi_{\mathrm{HO}},\psi_{\mathrm{AF}}]$ vs.~$\psi_{\mathrm{HO}}$ on the horizontal axis and $\psi_{\mathrm{AF}}$ on the vertical axis. The red arrows in Fig.~\ref{fig:LFC} indicate the state of the system during the described evolution. At point A for $T>T_{\mathrm{HO}}$, the system is at the energy minimum $\psi_{\mathrm{HO}} = \psi_{\mathrm{AF}} = 0$ as shown in Fig.~\ref{fig:LFC}(a). At point B for $T = 15.3$~K $< T_{\mathrm{HO}}$, the minimum at the origin splits into two degenerate minima on the horizontal axis shown in Fig.~\ref{fig:LFC}(b). Consequently, the system spontanelously breaks symmetry and acquires $\psi_{\mathrm{HO}} \ne 0$ via a second-order phase transition. Using the condition $\partial^2 f / \partial \psi_{\mathrm{HO}}^2 = 0$ at $\psi_{\mathrm{HO}} = \psi_{\mathrm{AF}} = 0$ for the transition temperature $T_{\mathrm{HO}} = 17.5$~K, the interaction constant $J^x_- = 2\Delta/\tanh(\Delta/T_{\mathrm{HO}})\approx 46$~K can be deduced\cite{haule2010complex}. At a lower temperature, such as $T=3$~K for point C, the free energy develops a second pair of shallower (local) minima along the vertical (magnetic) axis as shown in Fig.~\ref{fig:LFC}(c), but the system stays at one of the global minima with $\psi_{\mathrm{HO}} \ne 0$ and $\psi_{\mathrm{AF}} = 0$.  Under pressure, the AF minima on the vertical axis become deeper than the non-magnetic minima on the horizontal axis, so the system undergoes a first-order phase transition from HO to AF with $\psi_{\mathrm{AF}} \ne 0$ and $\psi_{\mathrm{HO}} = 0$ at high pressure \cite{haule2010complex}. To explain the first order of the phase transition, we choose a large enough $\gamma \approx 64$~K to ensure an energy barrier separating the minima on the magnetic and non-magnetic axes. This picture is supported by Raman spectroscopy \cite{PhysRevLett.117.227601} in Fe-doped URu$_2$Si$_2$, where optically-induced transitions between the HO and AF minima in the energy landscape were observed. Using the value $T_{\mathrm{AF}} = 15$~K extrapolated to ambient pressure \cite{haule2010complex} and its associated condition $\partial^2 f / \partial \psi_{\mathrm{AF}}^2 = 0$ at the origin, we deduce $J^y_- = 2\Delta/\tanh(\Delta/T_{\mathrm{HO}})\approx 43$~K.

\section{Competition of Hidden Order and Ferromagnetism}
Contributions to the PKE from alternating AF layers cancel out in the bulk, but the contribution from the surface layer may produce a non-zero PKE \cite{PhysRevLett.75.3004}. However, its sign cannot be trained by a uniform external magnetic field\cite{yakovenko2015tilted}, so the AF scenario is not a viable explanation for the experiment in Ref.~\onlinecite{PhysRevB.91.140506}. Thus, we turn our attention to non-staggered FM order $\psi_{\mathrm{FM}}$. The training magnetic field $B$ couples to it linearly in Eq.~\eqref{eqn:hamiltonian}, thus lowering the energy of the FM state and making it competitive with HO. In contrast, AF has higher energy than HO at ambient pressure, so we set $\psi_{\mathrm{AF}} = 0$, and the free energy per site in Eq.~\eqref{eqn:hklandau} becomes
\begin{eqnarray}
&&f[\psi_{\mathrm{HO}},\psi_{\mathrm{FM}}] = J^x_- \psi_{\mathrm{HO}}^2 + J^y_+ \psi_{\mathrm{FM}}^2+ \gamma\psi_{\mathrm{HO}}^2\psi_{\mathrm{FM}}^{2}  \label{eqn:IntFreeEnergy}\\
&&-T\ln\left[\cosh\left(\beta \sqrt{\Delta^2 + \left({J^x_- \psi_{\mathrm{HO}}}\right)^2+\left({J^y_+ \psi_{\mathrm{FM}} + b}\right)^2}\right)\right]. \nonumber
\end{eqnarray}
Eq.~\eqref{eqn:IntFreeEnergy} differs from Eq.~\eqref{eqn:AFFreeEnergy} by the coefficient $J^y_- \rightarrow J^y_+$ and the presence of magnetic energy $b$. The difference between $J^y_+$ and $J^y_-$ is only due to the small interlayer coupling $J^y_{\bot}$, so $J^y_+$ still has a positive sign favorable for FM. Since the value of $J^y_{\bot}$ is unknown, we take $J^y_+ \approx 43$~K as an estimate. The observation of a FM phase in Re, Tc, and Mn doped samples \cite{PhysRevB.39.2423, dalichaouch1990ferromagnetism, butch2010suppression, PhysRevLett.103.076404} indicates that FM can, indeed, be a close competitor of HO.

Let us compare two experimental protocols employed in Ref.~\onlinecite{PhysRevB.91.140506} for going from point A to point C in Fig.~\ref{fig:phase_diagram}: zero-field cooling (ZFC) via A-B-C and high-field cooling (HFC) via A-D-E-F-G-C. The energy landscape of Eq.~\eqref{eqn:IntFreeEnergy} at points A, B, and C is shown in Fig.~\ref{fig:LFC}(a)-(c) and has already been discussed below Eq.~\eqref{eqn:AFFreeEnergy}, but now the vertical axis represents $\psi_{\mathrm{FM}}$ instead of $\psi_{\mathrm{AF}}$. During ZFC, the system undergoes a second-order phase transition to the HO ground state with $\psi_{\mathrm{HO}} \ne 0$ and $\psi_{\mathrm{FM}} = 0$, and stays there as temperature decreases.

Now let us consider HFC starting at point A, where the energy minimum is located at $\psi_{\mathrm{HO}} = \psi_{\mathrm{FM}} = 0$ as shown in Fig.~\ref{fig:LFC}(a). Next, a training magnetic field $b=0.4$~K is applied (point D in Fig.~\ref{fig:phase_diagram}) shifting the energy minimum in the FM direction $\psi_{\mathrm{FM}}>0$ as shown in Fig.~\ref{fig:LFC}(d). At point E with $T = 15.3$~K, the free energy develops two shallow degenerate HO minima, but the system stays in the pre-existing FM global minimum as shown in Fig.~\ref{fig:LFC}(e). At nearby point F with $T=15$~K, the HO minima become deeper than the FM minimum as seen in Fig.~\ref{fig:LFC}(f), but the energy barriers prevent a transition. So, the system stays in the metastable FM minimum all the way down to $T=3$~K at point G, as shown in Fig.~\ref{fig:LFC}(g). Removing the magnetic field at $T=3$~K takes the system to point C in Fig.~\ref{fig:phase_diagram} while preserving its FM state as depicted in Fig.~\ref{fig:LFC}(h). Although the energy landscape in panel (h) is exactly the same as in panel (c), the state of the system is different: It is HO for ZFC and FM for HFC. The metastable FM state is reached because HFC crosses the first-order rather than the second-order phase transition line in Fig.~\ref{fig:phase_diagram}. Finally, when temperature is increased along the path C-B-A at $b=0$, the FM metastable state exhibits a non-zero PKE, as observed on warmup at zero field in Ref.~\onlinecite{PhysRevB.91.140506}.

The theoretical scenario presented above offers a qualitative explanation of experiment \cite{PhysRevB.91.140506} but has shortcomings. First, the experimental PKE persists on warmup to $T > T_{\mathrm{HO}}$, whereas in our model the FM minimum in free energy disappears at $T < T_{\mathrm{HO}}$. Second, the PKE magnitude observed in Ref.~\cite{PhysRevB.91.140506} increases with the increase of the training magnetic field. This feature can be explained theoretically by considering partial statistical population of different states in the energy landscape due to thermal fluctuations. However, further refinements of the model are beyond the scope of this paper and are left for future studies.

\section{Field-Reversal Test}
\begin{widetext}

\begin{figure*}[!h]

\includegraphics[width=1.0\textwidth]{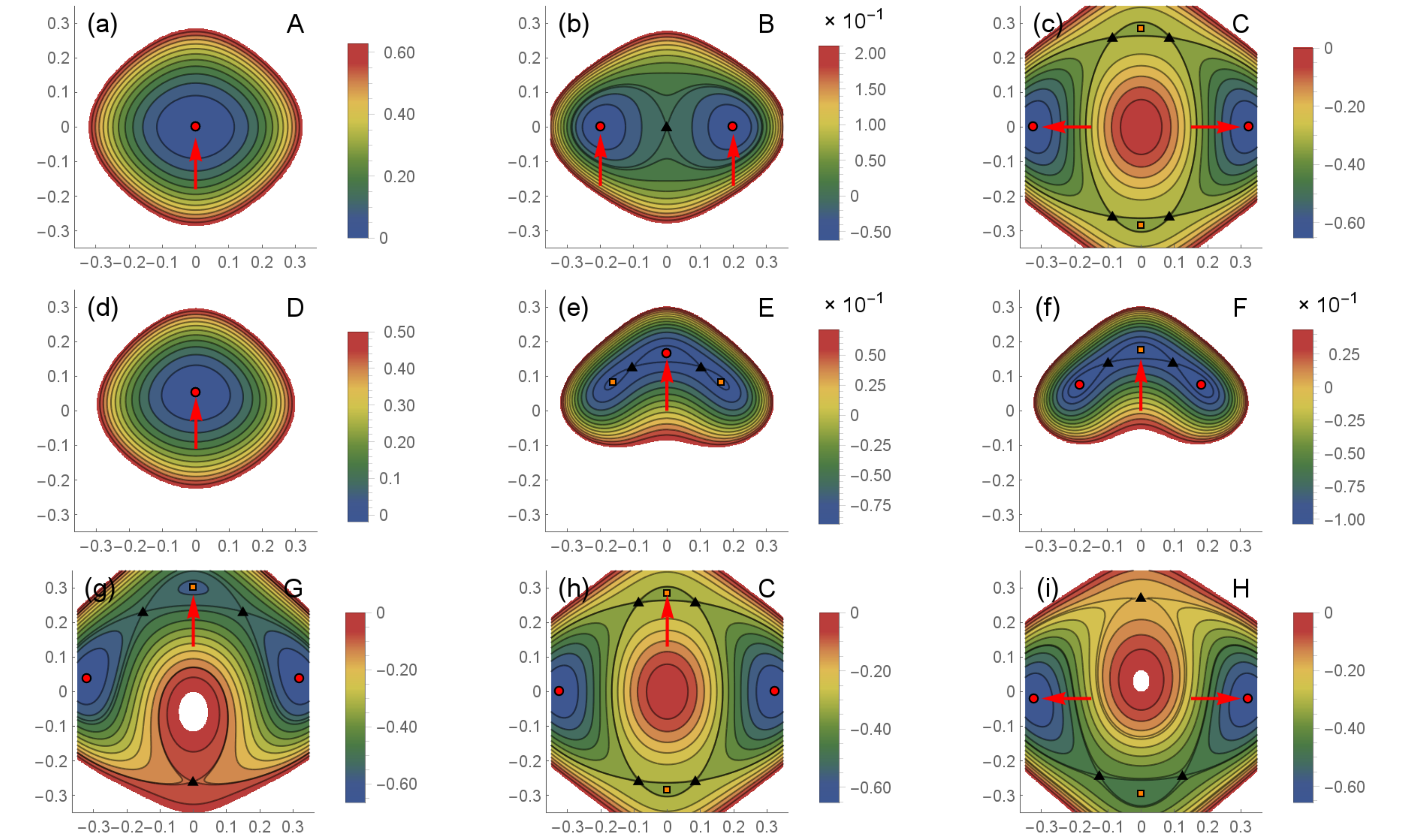}
\caption{
Contour plots of the free energy $f[\psi_{\mathrm{HO}},\psi_{\mathrm{AF}}]$ given by Eq.~\eqref{eqn:AFFreeEnergy} or $f[\psi_{\mathrm{HO}},\psi_{\mathrm{FM}}]$ given by Eq.~\eqref{eqn:IntFreeEnergy} for points A-H in Fig.~\ref{fig:phase_diagram}. The horizontal and vertical axes represent the non-magnetic, $\psi_{\mathrm{HO}}$, and magnetic, $\psi_{\mathrm{AF}}$ for (a)-(c) and $\psi_{\mathrm{FM}}$ for (a)-(i), components of the order parameter. Global minima, local minima, and saddle points are indicated by red disks, orange squares, and black triangles, while red arrows indicate the state of the system reached following the paths in Fig.~\ref{fig:phase_diagram}.
}\label{fig:LFC}
\end{figure*}

\end{widetext}

The proposed scenario can be tested by applying a reversed magnetic field, in the opposite direction relative to the HFC training field, at low temperature. When the magnetic energy reaches a critical magnitude $-b_1 \approx -0.22$~K corresponding to point H in Fig.~\ref{fig:phase_diagram}, the metastable FM minimum transforms into a saddle point as shown in Fig.~\ref{fig:LFC}(i), so the system makes an irreversible transition to one of the HO minima indicated by the red arrows. This transition can be detected by applying and removing a progressively increasing reversed magnetic field at low temperature, while measuring the PKE at $b=0$ in each cycle.

Instead of using the optical PKE technique, the metastable FM can also be observed by direct magnetization measurements\cite{PARK1998455} using a sensitive probe, such as a SQUID magnetometer. The magnetic moment in the FM state can be crudely estimated to be of the same order as the staggered magnetic moment $m_\mathrm{AF} = 0.3\mu_B$ experimentally measured\cite{PhysRevB.82.060408} in the AF phase. However, the magnetic moment in the metastable FM state would be greatly reduced by thermal fluctuations between the global and local minima in Fig.~\ref{fig:LFC}. Therefore, the effective FM moment is expected to be small, so that direct measurement of magnetization would require high sensitivity, consistent with the PKE sensitivity. The field-reversal test of the metastable FM state can also be performed using direct magnetization measurements.

\section{Comparison with Experiment}
The magnetic energy $b_1$ in the field-reversal test is one of the several characteristic magnetic energies $b_1$, $b_2$, $b_3$ shown in Fig.~\ref{fig:phase_diagram}, indicating qualitative changes in the free-energy landscape in Fig.~\ref{fig:LFC}. The magnetic energy $b_2$ corresponds to the first-order phase transition between $\psi_\mathrm{HO}$ and $\psi_\mathrm{FM}$, where the free energy $f_\mathrm{HO}$ of the HO minima in Fig.~\ref{fig:LFC} is equal to the free energy $f_\mathrm{FM}$ of the FM minimum. The magnetic energy $b_3$ corresponds to the termination of the metastable HO phase, where the HO minima in Fig.~\ref{fig:LFC} disappear. Experimentally, HO terminates at a magnetic field of about $35$~T \cite{correa2012high}. For comparison of theory with experiment, we need to convert magnetic energy $b$ in Kelvins into magnetic field $B$ in Teslas. The conversion coefficient can be estimated as $B/b = \mu^{-1}_{\mathrm{eff}} = 1.2$~T/K using the effective magnetic moment $\mu_{\mathrm{eff}} = |\bra{A_2}L_z+2S_z\ket{A_1}|\mu_B = 1.25\mu_B$ quoted in Ref.~\onlinecite{haule2010complex}. However, for $b_3 = 0.93$ K in Fig.~\ref{fig:phase_diagram}, this $\mu_{\mathrm{eff}}$ gives the terminating field $B_3 = 1.1$~T, which is far short of the $35$~T seen in experiment. This discrepancy can be resolved in two ways.

The value $b_3 = 0.93$~K shown in Fig.~\ref{fig:phase_diagram} was obtained for particular values of the unknown parameters $\Delta$, $J^y_+$, and $\gamma$ and can be increased by adjusting those parameters. A formula for $b_3$ is derived in Appendix A, and the maximal value $b^{(max)}_3 = T_\mathrm{HO}$ is achieved in the limit $\gamma \rightarrow \infty$ and $\Delta \rightarrow 0$. Using $\mu^{-1}_\mathrm{eff} = 1.2$~T/K and $b^{(max)}_3 = T_\mathrm{HO} = 17.5$~K, we obtain $B_3 = 21$~T, which is closer to the experimental value.

Moreover, the conversion coefficient $\mu_\mathrm{eff}$ can be estimated from experiment, rather than from the theoretical quote in Ref.~\onlinecite{haule2010complex}. The staggered moment observed in the antiferromagnetic phase in experiment\cite{PhysRevB.82.060408} is $m_\mathrm{AF} = 0.3\mu_B$ per uranium atom. Comparing with the theoretical formula in Eq.~\eqref{eqn:ApBMagAF1} in Appendix B, we find $\mu_\mathrm{eff} = 0.3\mu_B$ in the limit $\Delta \rightarrow 0$, which is four times lower than the prior estimate. Combining this estimate for $\mu_\mathrm{eff}$ with the estimate for the maximal $b^{(max)}_3 = T_\mathrm{HO} = 17.5$~K, we obtain $B_3 = 87$~T, which exceeds 35~T by a wide margin. It shows that the theoretical estimate of the HO terminating magnetic field can be made large enough to match experiment by tuning the parameters of the model.

\begin{figure}[t]
\includegraphics[width=0.47\textwidth]{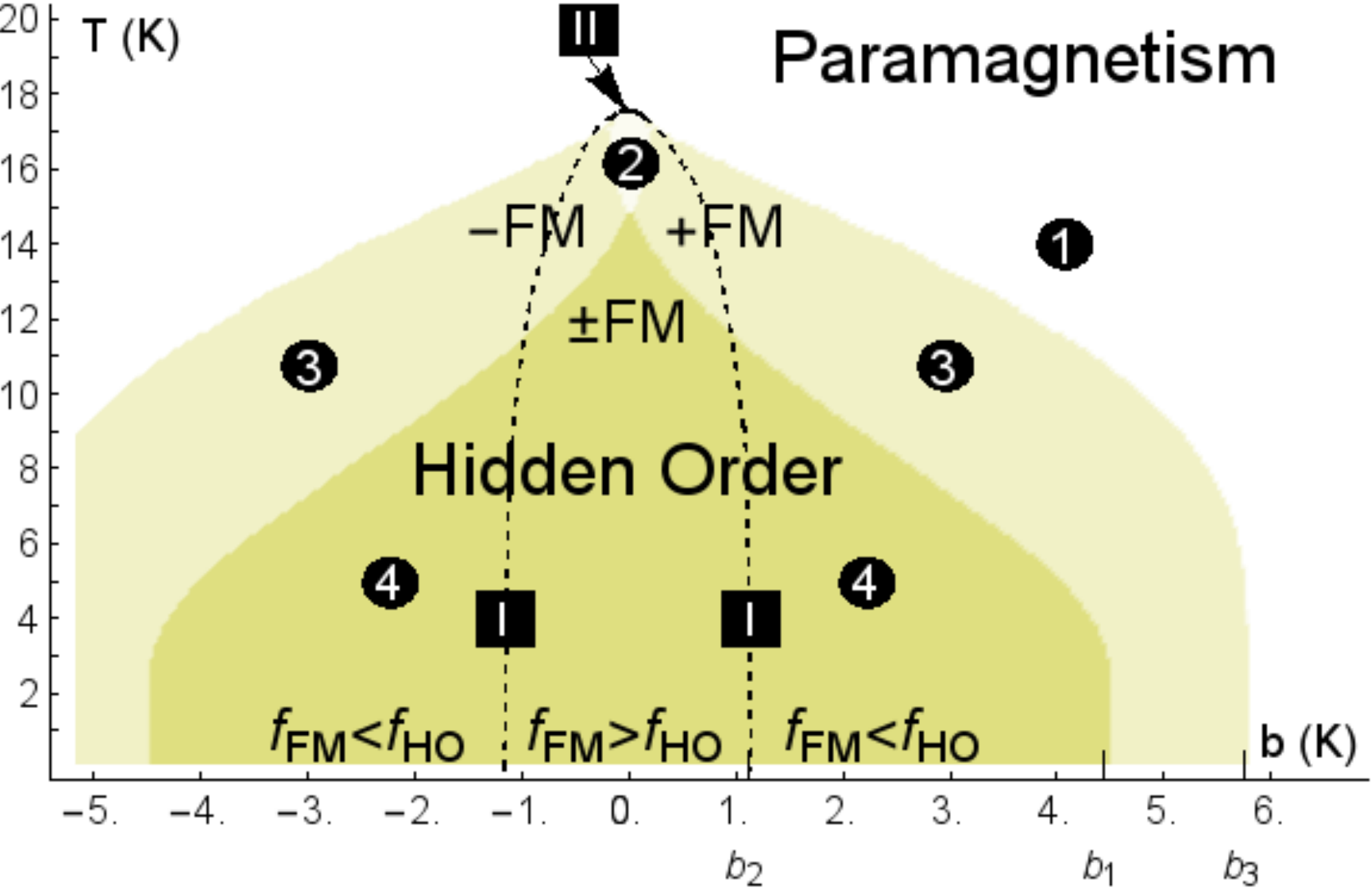}
\caption{Phase diagram as in Fig.~\ref{fig:phase_diagram} recalculated using $\Delta = 7$~K and $\gamma = 525$~K (in contrast to $\Delta = 17.5$~K and $\gamma = 64$~K in Fig.~\ref{fig:phase_diagram}). Notice the greater scale for the magnetic energy $b$.}\label{fig:phase_diagram2}
\end{figure}

For illustration we repeat the calculation for alternative values $\Delta = 7$~K and $\gamma = 525$~K and the corresponding generated values of $J^x_- = 2\Delta/\tanh(\Delta/T_\mathrm{HO}) = 37$~K, $J^y_+ = J^y_- = 2\Delta/\tanh(\Delta/T_\mathrm{AF})=32$~K, and $\mu_\mathrm{eff} \approx 0.33\mu_B$. The new phase diagram, shown in Fig.~\ref{fig:phase_diagram2}, shares qualitative features with Fig.~\ref{fig:phase_diagram}, but the characteristic energies $b_1$ and $b_2$ are interchanged. The HO termination energy $b_3=6$~K translates into $B_3=27$~T, and the field-reversal energy $b_1 = 4$~K translates into $B_1 = 18$~T.

So, there is a wide range of possible values for the characteristic fields $B_1$ and $B_3$ depending on the model parameters. However, the phase diagram of URu$_2$Si$_2$ in a strong magnetic field is complicated with multiple phase transitions \cite{PhysRevLett.91.256401,PhysRevLett.89.287201,PhysRevB.68.020406} not captured by our simple model. Additionally, the applicability of the Haule-Kotliar framework in very strong fields is not clear, as the basis states may change. So, our model should be primarily considered a qualitative, rather than quantitative, guide to experiment.

\section{Conclusions}
We have proposed a theoretical scenario reconciling the TR invariance of the HO state with observation of a non-zero magnetic-field-induced PKE \cite{PhysRevB.91.140506}. Competition between the real and imaginary parts of a complex order parameter in a generalized Haule-Kotliar model \cite{haule2010complex} results in either ground-state HO or metastable FM, depending on the path taken through the phase diagram. Our theory can be tested by applying a strong enough reversed magnetic field at low temperature, which should trigger a transition from FM to HO and cause the PKE to vanish. Although some issues remain open in our scenario, it has the advantage of giving a unified description of the HO and FM states within a single theoretical model without invoking extrinsic effects.

In principle, the general approach presented in our paper can be adapted to other two-level models of HO in the literature. In particular, the hastatic order proposed in Refs.~\onlinecite{chandra2013hastatic, PhysRevB.91.205103} is based on the 5f$^3$ configuration described by the effective spin $1/2$ and could also be used to explain intrinsic magnetism. However, the hastatic model predicts an in-plane magnetic moment in the HO phase which is not observed experimentally\cite{das2013absence,PhysRevB.89.155122}.

A non-zero PKE is also observed in the superconducting phase of URu$_2$Si$_2$ \cite{PhysRevB.91.140506} emerging from the HO phase below $T_c = 1.5$~K. A generalized model for the two separate TR symmetry breakings in the HO and superconducting phases, independently controllable by a training magnetic field \cite{PhysRevB.91.140506}, remains a challenge for future study.

\textit{This research did not receive any specific grant from funding agencies in the public, commercial, or not-for-profit sectors.}

\textit{Acknowledgments.} We thank G.~Blumberg, H.-H.~Kung, K.~Haule, N.~Butch, A.~Kapitulnik, and M.~Jaime for helpful discussions.

\bibliography{uru2si2}

\appendix
\section{The magnetic field terminating hidden order}
Here we evaluate the critical magnetic field energy $b_3$ corresponding to the termination of HO at $T=0$ on the phase diagrams shown in Figs.~\ref{fig:phase_diagram} and \ref{fig:phase_diagram2}. It can be derived from the free energy in Eq.~\eqref{eqn:IntFreeEnergy} at $T=0$,
\begin{eqnarray}
f[\psi_{\mathrm{HO}},\psi_{\mathrm{FM}}] &&= J^x_- \psi_{\mathrm{HO}}^2 + J^y_+ \psi_{\mathrm{FM}}^2+ \gamma\psi_{\mathrm{HO}}^2\psi_{\mathrm{FM}}^{2}\label{eqn:appFE}\\
&&-\sqrt{\Delta^2 + \left({J^x_- \psi_{\mathrm{HO}}}\right)^2+\left({J^y_+ \psi_{\mathrm{FM}} + b}\right)^2}. \nonumber
\end{eqnarray}
A general consideration is somewhat complicated, so we study the limiting cases of $\gamma = 0$ and $\gamma \rightarrow \infty$. 

The case of $\gamma = 0$ corresponds to the Haule-Kotliar model of Ref.~\onlinecite{haule2010complex}, but we arrive at a different result for $b_3$. At $\gamma = 0$, minimization of the free energy in Eq.~\eqref{eqn:appFE} gives two equations $\partial f/\partial \psi_{\mathrm{HO}}=\partial f/\partial \psi_{\mathrm{FM}}=0$:
\begin{eqnarray}
&&2\psi_{\mathrm{HO}}  = \frac{J^x_-\psi_{\mathrm{HO}}}{\sqrt{\Delta^2 + (J^x_- \psi_\mathrm{HO})^2 + (J^y_+ \psi_\mathrm{FM}+b)^2}} \label{eqn:ApAg0MF1},\\
&&2\psi_{\mathrm{FM}}  = \frac{J^y_+\psi_{\mathrm{FM}}+b}{\sqrt{\Delta^2 + (J^x_- \psi_\mathrm{HO})^2 + (J^y_+ \psi_\mathrm{FM}+b)^2}}.\label{eqn:ApAg0MF2}\end{eqnarray}
From Eq.~\eqref{eqn:ApAg0MF1} we find
\begin{eqnarray}
2\sqrt{\Delta^2 + (J^x_- \psi_\mathrm{HO})^2 + (J^y_+ \psi_\mathrm{FM}+b)^2} = J^x_-,\label{eqn:ApAg0MF4}
\end{eqnarray} 
and then from Eq.~\eqref{eqn:ApAg0MF2} we find $\psi_\mathrm{FM} = b/(J^x_- - J^y_+)$. The HO vanishes at the termination field $b = b_3$, where $\psi_\mathrm{HO} = 0$. Using these values for $\psi_\mathrm{HO}$ and $\psi_\mathrm{FM}$ in Eq.~\eqref{eqn:ApAg0MF4}, we find a formula for $b_3$:
\begin{eqnarray}
b_3(\gamma = 0) = \frac{J^x_- - J^y_+}{2}\sqrt{1-\left(\frac{2\Delta}{J^x_-}\right)^2}.\label{eqn:lowgammabc}
\end{eqnarray}
Eq.~\eqref{eqn:lowgammabc} replaces an incorrect formula on page 3 of Ref.~\onlinecite{haule2010complex} for the critical field $b_c$ corresponding to our $b_3$. The formula in Ref.~\onlinecite{haule2010complex} gives $b_c \propto J^x_- + J^x_+$, which cannot be valid, because a correct formula must give $b_3 \rightarrow 0$ in the limit $J^x_- \rightarrow J^y_+$, where an infinitesimal magnetic field would be necessary to favor FM over HO.

In the case $\gamma \rightarrow \infty$, the term $\gamma\psi_{\mathrm{HO}}^2\psi_{\mathrm{FM}}^{2}$ in Eq.~\eqref{eqn:appFE} imposes a high energy penalty for the co-existence of $\psi_\mathrm{HO}$ and $\psi_\mathrm{FM}$, so we set $\psi_\mathrm{FM} = 0$. Using this value and $\psi_\mathrm{HO}$ = 0 in Eq.~\eqref{eqn:ApAg0MF4}, we find
\begin{eqnarray}
b_3(\gamma \rightarrow \infty) = \frac{J^x_-}{2}\sqrt{1 - \left(\frac{2\Delta}{J^x_-}\right)^2} = \frac{\Delta}{\sinh(
\frac{\Delta}{T_\mathrm{HO}})},\label{eqn:highgammabc}
\end{eqnarray}
where the second equality follows from $2\Delta/J^x_- = \tanh(\Delta/T_\mathrm{HO})$.

Comparing Eqs.~\eqref{eqn:lowgammabc} and \eqref{eqn:highgammabc}, we observe that the highest termination field is achieved in our model in the limit $\gamma \rightarrow \infty$ and $\Delta \rightarrow 0$, where Eq.~\eqref{eqn:highgammabc} gives
\begin{eqnarray}
b_3^{(max)} = T_\mathrm{HO}.
\end{eqnarray}

\section{The Staggered Magnetic Moment}
Here we evaluate the staggered magnetic moment in the antiferromagnetic phase under pressure. We introduce a local magnetic field $b_j$, so that the free energy is given by Eq.~\eqref{eqn:hklandau} with $b \rightarrow b_j$. The local on-site magnetic moment $m_j$ at $b_j = 0$ and $T = 0$ is given by
\begin{eqnarray}
&&m_j = -\frac{\partial F}{\partial B_j} = -\mu_{\mathrm{eff}} \frac{\partial  F}{\partial b_j} = \\
&&= \mu_{\mathrm{eff}}\frac{\left({\sum_k J^y_{jk}\psi^y_k}\right)}{\sqrt{\Delta^2 + \left({\sum_k J^x_{jk}\psi^x_k}\right)^2+\left({\sum_k J^y_{jk}\psi^y_k}\right)^2}}.
\end{eqnarray}

In the antiferromagnetic phase, we have $\psi^y_n = (-1)^n\psi_{\mathrm{AF}}$ and $\psi^x_n =0$, so the staggered magnetic moment is $m_n = (-1)^n m_\mathrm{AF}$ where
\begin{eqnarray}
m_{\mathrm{AF}} = \mu_{\mathrm{eff}}\frac{J^y_- \psi_{\mathrm{AF}} }{\sqrt{\Delta^2 + (J^y_- \psi_{\mathrm{AF}})^2}.\label{eqn:ApBMagAF0} }
\end{eqnarray}
Using the minimum condition $\partial f / \partial \psi_\mathrm{AF} = 0$ for $f$ in Eq.~\eqref{eqn:AFFreeEnergy} at $T=0$, we find
\begin{eqnarray}
\psi_\mathrm{AF} = \sqrt{1-\left(\frac{2\Delta}{J^y_-}\right)^2}
\end{eqnarray}
and
\begin{eqnarray}
m_{\mathrm{AF}} = \mu_\mathrm{eff} \sqrt{1-\left(\frac{2\Delta}{J^y_-}\right)^2}=\frac{\mu_\mathrm{eff}}{\cosh(\Delta/T_\mathrm{AF}) }.\label{eqn:ApBMagAF1}
\end{eqnarray}
where we have used $2\Delta/J^y_- = \tanh(\Delta/T_\mathrm{AF})$.

The formula for the staggered magnetic moment $m_{(0,0,1)}$ given on page 3 of Ref.~\onlinecite{haule2010complex} differs from our Eq.~\eqref{eqn:ApBMagAF1} by an extra factor of 1/2, which we believe is incorrect.

\end{document}